# PyNMC: An Open-Source Framework for Neutron Multiplicity Counting Simulation Coupling OpenMC, FREYA, and ALPHANSO

Christopher Fichtlscherer[a,*]

[a]Department of Nuclear Science and Engineering,
Massachusetts Institute of Technology,
Cambridge, MA 02139, USA

*Corresponding author. Email: fcp@mit.edu*

**Abstract** – Neutron multiplicity counting (NMC) underpins plutonium assay in nuclear safeguards, arms control, and disarmament verification, but existing simulation tools are essentially limited to MCNPX-PoliMi [1] (export-controlled, MCNP license required) and ONMS [2] (open-source but built on Geant4 with no scripting API); other codes (RMC, MCNP-PTA) are institute-internal. We present PyNMC, an open-source, Python-native NMC simulation framework that couples OpenMC for transport with FREYA for event-by-event correlated prompt-neutron emission and ALPHANSO for native $(\alpha, n)$-source estimates, together with collision-level time-tagged event recording and a Python shift-register post-processor cross-validated against ONMS. The framework is validated against the ESARDA Neutron Multiplicity Benchmark on bare $^{252}$Cf (c2-10, c2-100), the low-multiplication Pu metal case c3s ($M = 1.12$ from an independent $k$-eigenvalue calculation; ESARDA spec $M = 1.08$), and a 10 g $PuO_2$ sample with an $(\alpha, n)$-source term (c4s); an internal stress-test extension to a $\approx 100$ g Pu metal sample at $M = 1.29$ is reported alongside but lies beyond the ESARDA participant range. For c4s, ALPHANSO gives $\alpha = 0.78$ with modern cross-section data; the reported benchmark comparison rescales the $(\alpha, n)$ rate to the ESARDA value $\alpha = 0.853$. Simulated rates agree with point-model predictions for all cases, and with the published ESARDA participant-code scatter where participant results exist. The framework is shipped as a Docker container under the MIT license and is openly available on GitHub at `github.com/cfichtlscherer/nmc`.

## 1. Introduction

Neutron multiplicity counting (NMC) is a non-destructive, non-intrusive assay technique that quantifies plutonium content by analysing time-correlated groups of neutrons from spontaneous fission [3]. Determination of full isotopic composition typically requires complementary information (gamma spectroscopy or declared isotopics). NMC is widely used in nuclear safeguards and arms control verification, where measurements must characterise plutonium content while respecting information barriers [4], and where simulation is required to quantify and correct systematic measurement-to-mass biases for warhead authentication [5].

The technique measures three rates: Singles ($S$), Doubles ($D$), and Triples ($T$). From these three observables, three unknown sample properties can be determined simultaneously: the spontaneous fission rate $F_s$ (proportional to $^{240}$Pu effective mass), the neutron multiplication $M$, and the $(\alpha, n)$-to-spontaneous-fission ratio $\alpha$ [6, 7]. The theoretical foundation is the point model [8], which relates the measured rates to sample parameters through the factorial moments of the fission multiplicity distribution $P(\nu)$.

Simulation tools that predict multiplicity counter response are needed for detector design, calibration transfer, and interpretation of measurements on complex items where the point model assumptions break down [9]. Monte Carlo particle transport can model full three-dimensional geometry, energy-dependent neutron transport, and detector response, and is particularly valuable for designing verification protocols and quantifying measurement sensitivity [2].

However, $S$, $D$, and $T$ are higher factorial moments of the detected neutron count – equivalently, two-time correlations of the detection process – and are not first moments of the neutron flux. They are therefore not recoverable from the usual first-moment linear Boltzmann flux solution alone, which yields only the expected phase-space density. Higher-moment generalisations exist – the Pál–Bell forward equation for the probability-generating functional [10, 11], and branching-process / deterministic moment methods more broadly – but are tractable only for idealised homogeneous geometries; the point model [8] is the zero-dimensional limit. Realistic counter geometries with heterogeneous moderators, energy-dependent transport, and multiplying samples require event-by-event Monte Carlo.

The established simulation tool in the safeguards community is MCNPX-PoliMi [1] (currently distributed as v2.0 by RSICC, with the fission-physics extensions of Pozzi *et al.* 2012 [12] adding $\nu$-dependent prompt neutron spectra and fragment-axis anisotropy on top of the original 2003 multiplet emission), which modifies MCNP [13] to emit fission neutrons as a correlated multiplet rather than as independent particles. Default MCNP samples each fission neutron's energy independently from the prompt fission spectrum and its direction isotropically, with no inter-neutron correlation; the MCNP 6.2 Release Notes are explicit about this, stating that "the MCNP library physics treatment suffers from uncorrelated secondary-particle production due to the sampling of inclusive data in the ACE libraries. While the average particle production is preserved, coincidence physics and scoring is not possible" and that the correlated event generators CGMF and FREYA were added in 6.2 as opt-in alternatives, activated only via the `FMULT` card with the `METHOD` keyword [14]. MCNPX-PoliMi is export-controlled and requires an MCNP license, hindering the transparent, reproducible simulation work needed for arms control. Another modified-MCNP NMC package, MCNP-PTA (JRC Ispra) [15], includes built-in shift-register electronics models but is restricted in distribution. A more recent code, RMC (Tsinghua), integrates the FREYA and CGMF correlated fission event generators into its own transport kernel [16] and has been applied to NMC simulation [17]; the source is not publicly distributed. The only fully open-source predecessor is ONMS [2], a Geant4-based simulation that uses the LLNL Fission Library [18]. ONMS demonstrated the viability of open-source NMC simulation and informed the design of the present work. It is built on Geant4 10.2 and uses a C++ codebase without a scripting API. Table 1 summarises the simulation-tool landscape.

| Tool | Transport kernel | Code availability | Correlated fission | $(\alpha, n)$-source | Scriptable workflow | Shift-register / pulse-train |
|---|---|---|---|---|---|---|
| MCNPX-PoliMi v2.0 (Politecnico di Milano) [1, 12] | MCNP | RSICC; export-controlled; requires MCNP license | Correlated multiplet, $\nu$-dep. spectra, fragment-axis anisotropy | Built-in ($PuO_2$, AmBe, AmLi) | Input deck | External (MPPost, bundled) |
| MCNP-PTA (JRC Ispra) [15] | MCNP | JRC-internal; requires MCNP license | $P(\nu)$ correlated | Built-in | Input deck | Built-in |
| RMC (Tsinghua University) [17] | RMC | No public distribution | Event-by-event correlated $\nu$, energies, angles (FREYA, CGMF) | Not generally available | Not generally available | Not generally available |
| ONMS (TU Darmstadt) [2] | Geant4 | Open (GPL); C++ source on GitHub | $P(\nu)$ correlated, independent kinematics | Built-in | C++; no scripting API | Built-in |
| **PyNMC** This work (MIT) | OpenMC (provided fork) | Open (MIT); Dockerfile on GitHub | Event-by-event correlated $\nu$, energies, angles (FREYA) | Built-in (ALPHANSO) | Python-native | Built-in |

**Table 1:** Open-source / scriptable feature matrix of the NMC simulation tool landscape. The "Correlated fission" column ranks tools by event-level fidelity, from $P(\nu)$-correlated multiplet emission (count right, kinematics inclusive) through MCNPX-PoliMi v2.0′s $\nu$-dependent spectra and fragment-axis anisotropy to event-by-event correlated prompt-neutron multiplicities, energies and angles via fragment-evaporation generators (FREYA, CGMF). "Not generally available" denotes a feature not documented in the public literature.

**Contributions.** The main contributions of this work are: (i) an OpenMC-compatible, time-tagged source and detection workflow for neutron multiplicity counting; (ii) coupling of FREYA to OpenMC for event-by-event spontaneous and induced fission with the full $P(\nu)$ distribution; (iii) integration of ALPHANSO for $(\alpha, n)$-source terms in plutonium oxide; (iv) a Python shift-register post-processor cross-validated against ONMS; (v) identification of a sampling-with-replacement bias in OpenMC's default `FileSource` that affects time-correlated calculations, with an opt-in sequential-iteration mode implemented in the provided OpenMC fork (planned for upstream submission); (vi) geometry-aware position rejection at source-write time against an arbitrary union of OpenMC `Cell` or `Region` domains, mirroring the role of OpenMC's `CoincidentSource constraints={"domains": [...]}` while keeping the existing `FileSource` pipeline intact[1] and (vii) validation against the ESARDA bare-source, low-multiplication, and $(\alpha, n)$-active benchmark cases. The framework is distributed under the MIT license with a Dockerfile that reproduces all simulation results of this paper end-to-end.

## 1.1. OpenMC as a Platform for NMC Simulation

OpenMC [19] is an open-source Monte Carlo particle transport code originally developed at MIT, with an active developer base, continuous integration, and extensive documentation. Several features make it suitable as a platform for multiplicity counting simulation:

OpenMC's transport kernel supports shared-memory parallelism, the full simulation workflow – geometry, materials, source, tallies, and post-processing – is scripted in Python, and user-defined sources, custom tallies, and collision-level event tracking provide the hooks needed for NMC simulation. As open-source software (MIT license), it removes the licensing and export-control barriers compared to MCNPX-PoliMi. Compared to ONMS it replaces a Geant4-based stack written in C++ with no scripting interface by an OpenMC-based stack with a Python

[1] A worked example for a non-convex source region ships as `examples/07_geometry_rejection.py` in the released code.

API for scripted geometry, source generation, transport, and post-processing, and delivers a per-core speedup of more than an order of magnitude (quantified in Section 2.2).

# 2. Modifications for Multiplicity Counting Simulation

Three modifications to the standard OpenMC simulation workflow are required for neutron multiplicity counting: (1) replacing the default fission multiplicity sampling with the correct $P(\nu)$ distribution, (2) generating time-correlated source particles and recording detection events with physical timing, and (3) implementing shift register analysis to extract $S$, $D$, and $T$ from the detection pulse train. Figure 1 shows the complete simulation chain; this section describes each component.

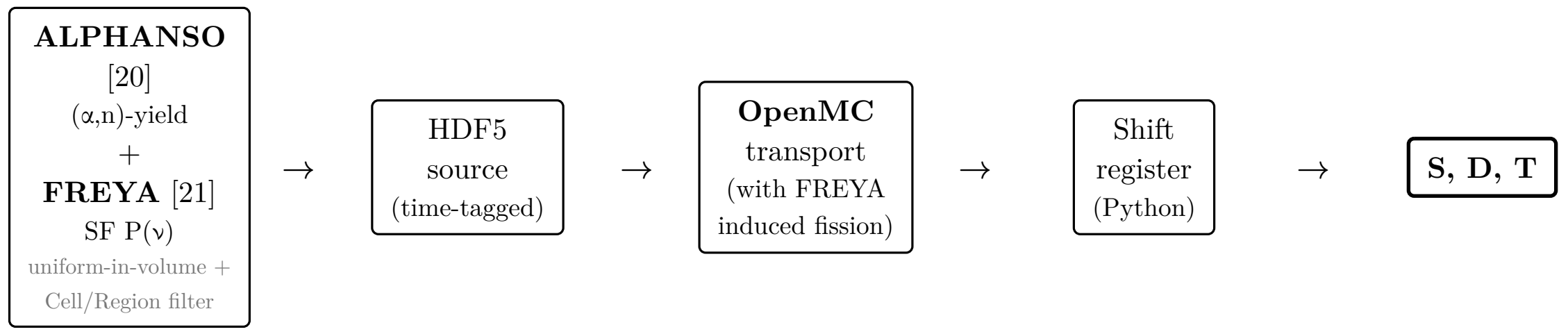

**Figure 1:** Simulation pipeline with three explicitly delineated stages. Source generation (left): ALPHANSO computes the $(\alpha, n)$-source rate and spectrum, and FREYA generates correlated spontaneous-fission multiplets; both are merged into a time-tagged HDF5 source file with one row per emitted neutron (per-event birth times shared within a fission, Poisson-distributed across fissions). Transport (centre): OpenMC reads the source file in sequential `FileSource` mode and transports neutrons through the AWCC detector, calling FREYA at each induced-fission site. Post-processing (right): $^3\mathrm{He}(n,p)^3\mathrm{H}$ absorption events are written to an HDF5 collision file and fed to a Python shift register that yields $S$, $D$, $T$.

## 2.1. The Multiplicity Sampling Problem

The factorial moments of the spontaneous fission multiplicity distribution $P(\nu)$ are defined as

$$\begin{aligned} \overline{\nu} = \nu_{s1} &= \sum_n n\,P(n), \\ \nu_{s2} &= \langle \nu(\nu-1) \rangle = \sum_n n(n-1)\,P(n), \\ \nu_{s3} &= \langle \nu(\nu-1)(\nu-2) \rangle = \sum_n n(n-1)(n-2)\,P(n), \end{aligned} \tag{1}$$

and map directly onto the NMC observables (Singles $\propto \nu_{s1}$, Doubles $\propto \nu_{s2}/2$, Triples $\propto \nu_{s3}/6$).

Standard OpenMC samples the number of fission neutrons from a two-point distribution on $\{\lfloor\overline{\nu}\rfloor, \lceil\overline{\nu}\rceil\}$ with weights chosen to preserve $\langle\nu\rangle = \overline{\nu}$. The second factorial moment of the two-point distribution then has the closed form

$$\langle \nu(\nu-1) \rangle_{\text{2pt}} = \lfloor\overline{\nu}\rfloor\,(2\overline{\nu} - \lfloor\overline{\nu}\rfloor - 1). \tag{2}$$

For $^{252}$Cf at FREYA's $\overline{\nu} = 3.775$ this evaluates to $3 \cdot (2 \cdot 3.775 - 3 - 1) = 10.65$, against the physical value $\nu_{s2} = 12.07$ from the Holden–Zucker distribution [22] – a 12% deficit in the second factorial moment that propagates directly to a comparable error in the Doubles rate, with an even larger error in the Triples (Figure 2).

To quantify the impact in a controlled setting, we compare two synthetic simulations of $5 \times 10^4$ $^{252}$Cf fission events at 1000 SF/s with $\varepsilon = 0.30$ and $\tau = 43$ $\mu$s. The only difference is the source multiplicity sampling: FREYA samples each event's $\nu$ from the full $P(\nu)$ distribution; the floor/ceil scheme samples $\nu \in \{3, 4\}$ with probabilities preserving $\overline{\nu} = 3.775$. The shift-register rates (Table 2) confirm the analytic prediction: the Singles rate is unaffected (both schemes preserve $\overline{\nu}$), the Doubles drops by 12.8% under floor/ceil – broadly consistent with the analytic $\nu_{s2}$ deficit of 11.8%, with the residual reflecting higher-order shift-register corrections beyond the leading $\nu_{s2}$ scaling and finite-statistics noise; and the Triples drops by 39%, dominated by the much larger $\nu_{s3}$ deficit.

| Rate | FREYA | Floor/ceil $\overline{\nu}$ | Ratio |
|---|---|---|---|
| $S$ (1/s) | 1129.3 | 1126.5 | 1.002 |
| $D$ (1/s) | 374.0 | 326.1 | 1.147 |
| $T$ (1/s) | 67.1 | 40.7 | 1.649 |

**Table 2:** Comparison of shift-register rates using FREYA $P(\nu)$ vs. floor/ceil $\overline{\nu}$ sampling, in two otherwise-identical synthetic simulations (same random seed, same detector parameters). The floor/ceil scheme reproduces the Singles rate to within 0.1% but yields $D$ and $T$ that are 13% and 39% lower than FREYA, because the two-point $\nu$-distribution does not match the true higher factorial moments.

This sampling is adequate for reactor physics calculations, where only the mean number of neutrons per fission matters for criticality. But for multiplicity counting, the variance and skewness of $P(\nu)$ carry the signal. The floor/ceil scheme is therefore not a physics-fidelity issue – OpenMC's underlying nuclear data, transport, and average multiplicity are unchanged – but a sampling-strategy issue: NMC requires the full $P(\nu)$ because the observables themselves are factorial moments beyond the mean.

## 2.2. Coupling FREYA for Correct $P(\nu)$ Sampling

**Choice of fission event generator.** Three open-source fission models are available for NMC simulation: FREYA [21], CGMF [23], and the LLNL Fission Library [18] used by ONMS [2]. FREYA and CGMF are physics-based event generators that emit a correlated multiplet – fragment partition is sampled, fragment excitation is set by energy conservation, and the neutron and prompt-gamma cascade is followed event by event, so the per-event ν, energies and angles are physically correlated. The LLNL Fission Library, by contrast, samples each emitted neutron's energy independently from a Watt spectrum and its direction isotropically with no inter-neutron correlation; this is what ONMS uses by default (Kütt notes that ONMS "does not rely on such correlation, mainly due to the strong moderation that takes place before neutron detection" [2]). For the integral observables relevant here – the first three factorial moments of $P(\nu)$ and the neutron energy spectrum – FREYA and CGMF agree at the few-percent level, comparable to or smaller than the inter-participant scatter visible in the ESARDA benchmark [24, 25]. We selected FREYA because it ships as a stand-alone shared library with a stable C interface and a smaller runtime footprint suitable for shift-register pulse-train rates. The coupling pattern – a Python source generator that writes a timed HDF5 source plus an in-transport hook for induced fission – applies unchanged with CGMF as the back end.

We replace the default multiplicity sampling by coupling FREYA version 2.0.3. (The ctypes-based Python interface used here accesses FREYA's Fortran `genfissevt_` entry point.) For each fission, FREYA:

- Samples fragment masses and kinetic energies from a physically motivated model
- Simulates sequential neutron and gamma emission from each fragment
- Produces correlated multiplicities, energies, and angular distributions
- Conserves energy and momentum

FREYA supports spontaneous fission of $^{238}$U, $^{238}$Pu, $^{240}$Pu, $^{242}$Pu, $^{244}$Cm, and $^{252}$Cf, and neutron-induced fission of $^{233}$U, $^{235}$U, $^{238}$U, $^{239}$Pu, $^{240}$Pu, and $^{241}$Pu. We use FREYA for both the initial spontaneous fission source and for induced fission during transport. A quantitative comparison against a multiplicity-only-correlated baseline (LLNL Fission Library / Watt spectrum) within our framework is left for future work.

A practical consequence of using FREYA inside OpenMC's transport loop is thread safety: FREYA carries global Fortran state and is not re-entrant, so all calls during transport are serialised via an OpenMP critical section. The section is entered only at induced-fission sites (a small fraction of all collisions) and the overhead is modest for non-multiplying samples but becomes significant for high-multiplication c3s and c4s runs. We measured OpenMP scaling at fixed problem size: 2.7 × at 8 threads (≈ 34% parallel efficiency) for non-multiplying c2-10, dropping to 1.1-1.2 × for the multiplying c3s case where FREYA serialisation dominates. Scaling out via independent realisations is therefore the more efficient strategy at the present problem size.

Measured wall-clock times on the reference host[2]: a smoke-test single c2-10 realisation (default `docker run --rm nmc`, OpenMC auto-thread) completes in 14 s; an isolated single c2-100 realisation (single-thread) in 4 min 03 s; an isolated single c4s realisation (single-thread) in 19 s[3]; the bare-source c2 validation campaign (twelve realisations of c2-10 and c2-100 combined, six each) in 27 min 00 s; and the full paper validation suite[4] (33 realisations across all five cases, single-threaded for determinism[5]) in 30 min 56 s (run_validation.py 27 min 00 s + run_pu_metal.py 2 min 24 s + run_pu_oxide.py 1 min 32 s).

**Comparison to ONMS.** Kütt reports that ONMS achieves "about 300 000 source events per hour computer time" [2], single-threaded, with the calculations in [2] carried out on Princeton's "Della" HPC cluster because ONMS supports neither native multithreading nor MPI. A single c2-100 realisation here (430 000 source fissions) completes single-threaded in 4 min 03 s on the reference host (see footnote[2]), or approximately $6.4 \times 10^6$ source fissions per hour, a factor of about twenty above the ONMS figure. Even after correcting for the decade-wide single-thread hardware-generation gap between the two reference platforms, a per-core advantage of well over an order of magnitude remains. Kütt himself identifies the structural reason as Geant4′s NeutronHP / ParticleHP performing on-the-fly Doppler broadening of neutron cross sections [2], where OpenMC uses pre-broadened pointwise ENDF data. On a multi-core host the per-core advantage compounds further with OpenMP threading (≈ 2.7 × at 8 threads on c2-10); this parallelism mode is not available in ONMS.

The three distributions compared in Figure 2 have distinct provenances. Holden–Zucker [22] is an empirical tabulation of $^{252}$Cf $P(\nu)$ assembled from direct neutron-emission measurements;

[2]All timings reported in this paper are wall-clock measurements on the same machine: AMD Ryzen 7 PRO 4750U (8 cores, 16 threads, base 1.7 GHz), 30 GB RAM, Ubuntu 24.04.4 LTS (Noble Numbat) with kernel 6.17, Docker 29.1.3. Measurements were taken on mains power; battery operation throttles single-thread workloads on this CPU by roughly 2–3 ×.

[3]c4s and the multiplying c3s and 100 g cases are run single-threaded because FREYA's serialised induced-fission calls dominate OpenMP scaling above ≈ 1.2 ×. The validation scripts also pin `OMP_NUM_THREADS = 1` for the bare-source c2-10 / c2-100 cases for run-to-run determinism, since OpenMP non-determinism otherwise leaks small (sub-permille) shifts into the absolute $S$, $D$, $T$ across reruns.

[4]Reproduces all rows of Table 3 and Table 5: six realisations each of c2-10 (8 600 SF/s), c2-100 (86 000 SF/s), the 100 g Pu extension ($M = 1.29$), and c4s ($\alpha = 0.853$); nine realisations of c3s ($M = 1.12$). 33 realisations in total.

[5]Single-threaded execution is deterministic given the seeds listed in the validation scripts, but the absolute simulated numbers depend on FREYA's Fortran-side `RANDOM_NUMBER` stream, which is fixed by the linked `libgfortran` at compile time. Different `gfortran` releases ship different default RNG algorithms (`KISS64` in `gcc` ≤ 6, `xoshiro256**` in `gcc` ≥ 7), so bit-for-bit reproducibility across rebuilds or hosts is not claimed; only statistical reproducibility within Monte Carlo scatter.

it is the canonical reference against which fission-model predictions are benchmarked. FREYA is a physics-based event generator in which $P(\nu)$ is an emergent output, not a fitted input: each fission samples a fragment partition $(A_1, A_2, TKE)$ – pre-neutron fragment masses and total kinetic energy – from measured pre-neutron mass and total-kinetic-energy yields, partitions the resulting fragment excitation by energy conservation, and lets a sequential Weisskopf evaporation cascade [26] run to completion – $\nu$ is whatever number of neutrons that cascade emits before each fragment falls below its neutron separation energy. FREYA's small set of free parameters ($e_0$, $x$, $c$, $c_S$, $d_{\text{TKE}}$) is tuned against $\overline{\nu}$, the average emission spectrum $\langle E_n \rangle$, and the sawtooth $\nu(A)$, but not against the shape of $P(\nu)$ itself or its higher factorial moments [21]. Therefore, agreement with Holden–Zucker on $\nu_{s2}$ and $\nu_{s3}$ is a non-trivial test of the fragment-evaporation physics, not a built-in feature. Floor/ceil, by contrast, is a delta distribution on the two integers bracketing $\overline{\nu}$ with weights chosen to preserve $\langle \nu \rangle$ and nothing else.

Numerically, FREYA reproduces $\nu_{s2}$ within 0.1% and $\nu_{s3}$ within 0.6% of the Holden–Zucker tabulation, while the floor/ceil two-point distribution underestimates $\nu_{s2}$ by 12% and $\nu_{s3}$ by 38% (figure caption). FREYA's $\overline{\nu} = 3.775$ is 0.5% higher than the ESARDA benchmark value of 3.756.[6] We use FREYA's value consistently throughout the simulation chain and the point-model comparison, so this difference cancels in validation ratios such as $S/S_{\text{pm}}$.

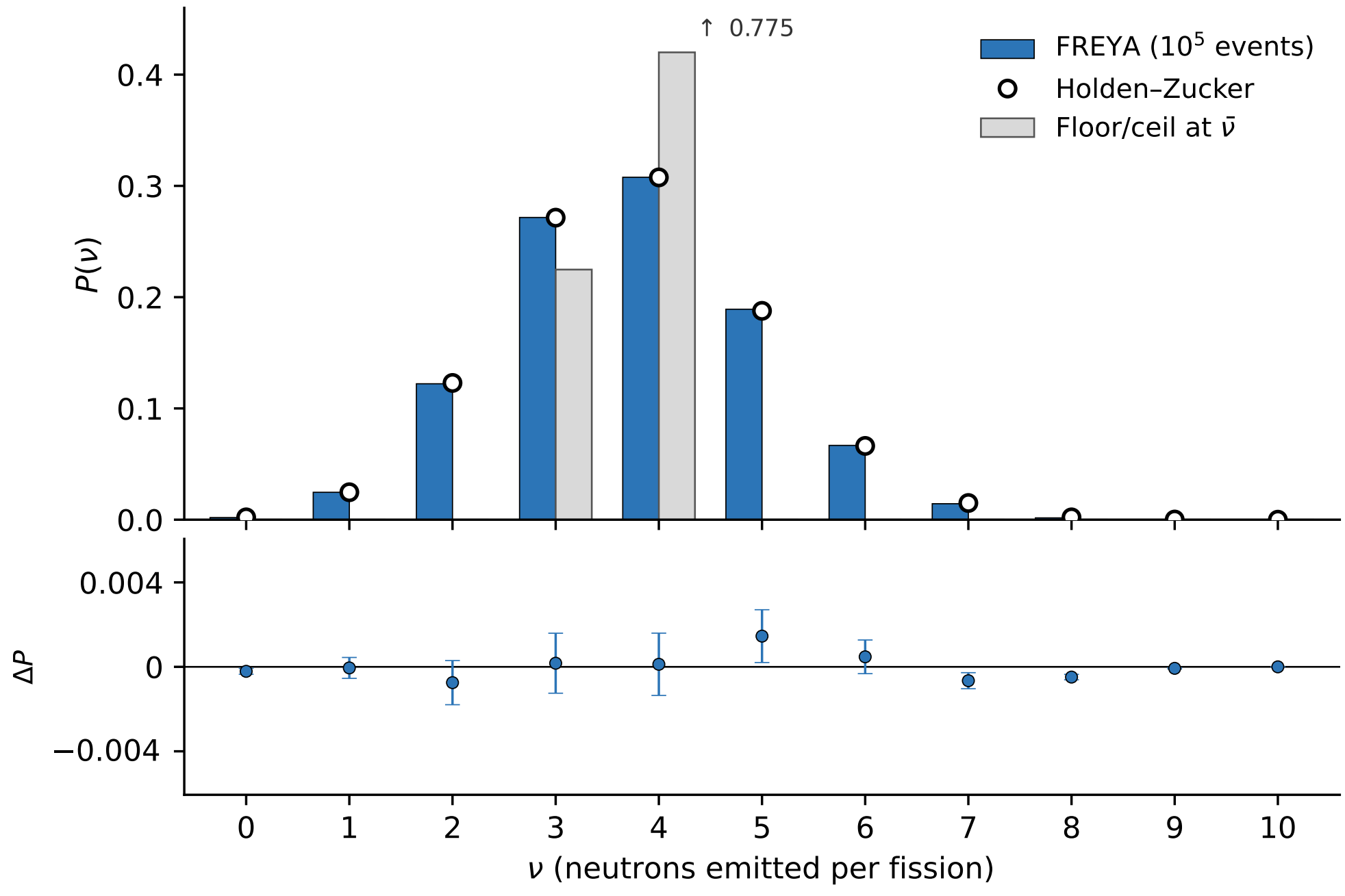

**Figure 2:** $^{252}$Cf spontaneous fission multiplicity distribution $P(\nu)$. Top panel: filled blue bars are FREYA ($10^5$ events); open black circles are the Holden–Zucker tabulation [22] (markers land on top of the FREYA bars across all $\nu$); hatched grey bars at $\nu = 3, 4$ are the standard $\overline{\nu} = 3.775$ floor/ceil two-point distribution, with the $P(4) = 0.775$ entry truncated for clarity at 0.42. Bottom panel: residual $\Delta P = P_{\text{FREYA}} - P_{\text{H-Z}}$ at each $\nu$ with $1\sigma$ binomial error bars from the $10^5$ FREYA samples; deviations are consistent with statistical noise. Factorial moments $(\nu_{s1}, \nu_{s2}, \nu_{s3})$: FREYA $(3.775, 12.05, 31.98)$, Holden–Zucker $(3.775, 12.07, 32.16)$, floor/ceil $(3.775, 10.65, 19.95)$. The floor/ceil scheme preserves $\overline{\nu}$ but not the variance or skewness of $P(\nu)$, and therefore cannot reproduce the higher factorial moments $\nu_{s2}$ and $\nu_{s3}$ that enter $D$ and $T$.

[6] The ESARDA benchmark [24] does not state $\overline{\nu}$ explicitly. Its value is implicit in Section 3.2 of [24], which pairs a $^{252}$Cf source intensity of 32 300 n/s with a fission rate of 8 600 SF/s for case 2a; the ratio 32 300 / 8 600 = 3.7558 rounds to $\overline{\nu} = 3.756$.

The coupling operates at two levels:

**Source generation.** For the initial spontaneous fission source, we generate fission events using FREYA's Python interface (via ctypes) and write them as an OpenMC source file. Each event produces $\nu$ neutrons (sampled from the full $P(\nu)$) with FREYA-determined energies and directions.

**Induced fission.** OpenMC's C++ fission handling has been modified to call FREYA at each induced fission site during transport; this coupling is validated for the c3s and 100 g Pu metal cases in Section 3.1.4.1.

## 2.3. Time-Correlated Event Recording

Multiplicity counting fundamentally depends on the time structure of detected events. Standard OpenMC simulations are stationary: particles have no absolute time stamps, and results are accumulated over batches. For NMC simulation, two timing-related modifications are needed.

First, source particles must carry physically meaningful birth times and a shared birth position. We assign times to spontaneous fission events by sampling from a Poisson process at the known source activity rate, and assign each event a single birth position sampled from the sample volume (a fixed point for $^{252}$Cf, uniform within the cylinder for the c3s Pu metal pellet and the c4s $PuO_2$ powder). All $\nu$ neutrons from a FREYA event share the same birth time and birth position. For samples with multi-region source distributions, the source-generation API accepts an optional `domains` argument (a list of OpenMC `Cell` or `Region` objects); candidate positions returned by the user-supplied sampler are then accepted only if they lie inside the union of the supplied domains, and rejected positions are resampled before the FREYA multiplet is written. This mirrors what OpenMC's `CoincidentSource` with `constraints={"domains": [...]}` does on the transport side.

Second, detection events must be recorded with their arrival times. We use a collision-tracking feature added to OpenMC in the provided fork, which records every interaction in user-specified cells to an HDF5 file, including the particle time, position, energy deposited, reaction type, and parent particle ID. Neutron detection in $^{3}$He proportional counters is identified by filtering for absorption events in the active gas volumes. The collision-track output records absorption with the ENDF sum label MT=101 (total neutron disappearance, the sum of all $(n, charged)$ and $(n, \gamma)$ channels); for $^{3}$He this sum is overwhelmingly the exoergic $(n, p)^{3}$H reaction (MT=103, $Q = 0.764$ MeV), which is the operative detection channel in real proportional counters. Filtering on MT=101 in $^{3}$He cells therefore captures every detection; an MT=103-only filter would yield the same pulse train to within sub-permille for $^{3}$He but is not used because the framework relies on the MT=101 disappearance label that OpenMC writes by convention.

A practical requirement: the number of simulated source particles must exactly equal the source-file size, otherwise OpenMC will either recycle source particles (creating duplicate detection times that inflate the coincidence rates) or leave entries on the floor. The framework provides a one-line helper `configure_filesource(settings, source_path)` that opens the HDF5 source bank, sets `settings.source`, `settings.particles = len(source)`, and `settings.batches = 1` consistently, so the invariant is enforced at the API level rather than left to the user.

## 2.4. Source-Bank Sampling for Time-Correlated Transport

OpenMC's default `FileSource` selects each transported particle by drawing a uniform random index into the source-bank file. Under the convention `settings.particles == len(source)` this is sampling with replacement: each source neutron is transported a Poisson-distributed number of times – some twice, some not at all, with mean one per neutron. For criticality and shielding, where only first moments of the flux matter, the effect averages out and goes unnoticed. For NMC it does not.

When the same source neutron is transported twice, the two transports share the source-event birth time exactly (it is recorded once in the HDF5 bank), but acquire independent transport histories – different scattering paths, different detection times. The shift register sees two detection events that share a per-fission timestamp and registers them as a real correlated pair, even though they originate from one source neutron transported twice rather than from two physically distinct neutrons emitted in the same fission.

We implement an opt-in sequential-iteration mode in OpenMC's `FileSource::sample()` in the provided OpenMC fork, activated by setting `OPENMC_FILESOURCE_SEQUENTIAL=1`. In this mode the source-bank sites are returned in order via an `std::atomic<size_t>` counter, so each source particle is transported exactly once when `settings.particles == len(source)`. Default OpenMC behaviour is preserved when the variable is unset, so users who want the standard with-replacement sampling for non-time-correlated calculations are unaffected. The atomic counter is shared across OpenMP threads within a single process, so OpenMP multi-thread parallelism is fully preserved – the OpenMP scaling reported earlier in this section is compatible with the sequential mode. MPI multi-rank distribution is not supported as-is, since each MPI rank has its own process memory; an MPI extension that partitions the source bank into rank-local contiguous slices is not exercised here.

### 2.5. Shift Register Post-Processing

The final component is a Python shift register that extracts $S$, $D$, and $T$ from the recorded detection pulse train. The implementation follows the standard multiplicity analysis algorithm [3]. For each detected neutron (trigger):

1. After a predelay $t_{\mathrm{pd}}$, a gate of width $t_g$ is opened (R+A gate)
2. The number of additional detections in this gate is recorded
3. A second gate at long delay $t_{\mathrm{ld}}$ records accidental coincidences (A gate)

Let $f_k$ and $b_k$ denote the $k$-th factorial moments of the R+A and A multiplicity distributions; the Singles, Doubles, and Triples rates are [3]:

$$S = \frac{N}{T_{\mathrm{meas}}} \tag{3}$$

$$D = S \cdot (f_1 - b_1) \tag{4}$$

$$T = \frac{S}{2} \cdot (f_2 - b_2 - 2b_1(f_1 - b_1)) \tag{5}$$

where $N$ is the total number of detections and $T_{\mathrm{meas}}$ is the length of the measurement window in seconds, so $S$ is the singles count rate. Eqs. (4)-(6) correspond to LA-13422-M [3] Eqs. 5-14 to 5-16. Triggers whose accidentals gate extends beyond the measurement window are excluded to avoid systematic bias in the background subtraction.

## 3. Validation

We validate the simulation chain on the ESARDA Neutron Multiplicity Benchmark, beginning with the bare $^{252}$Cf cases (c2-10, c2-100), with cross-code validation of the shift register against ONMS and a per-history comparison to the published ESARDA participants. Multiplying samples (c3s, $M = 1.12$, and a $\approx 100$ g extension at $M = 1.29$) and the $(\alpha, n)$-active $PuO_2$ case c4s (spec $\alpha = 0.853$) are addressed in the same section. Singles rates are compared to the point model; Doubles and Triples are validated via cross-code comparison and per-history factorial-moment analysis.

The reference against which we validate is the point model [3, 8]. The generalised expressions, accounting for leakage neutron multiplication $M$ and the $(\alpha, n)$-to-spontaneous-fission ratio $\alpha$, are LA-13422-M [3] Eqs. 5-44 to 5-46:

$$S_{\text{pm}} = F_s \varepsilon M \nu_{s1} (1 + \alpha) \tag{6}$$

$$D_{\text{pm}} = \frac{F_s \varepsilon^2 f_d M^2}{2} \left[ \nu_{s2} + \frac{M-1}{\nu_{i1}-1} \nu_{s1} \nu_{i2} (1+\alpha) \right] \tag{7}$$

$$T_{\text{pm}} = \frac{F_s \varepsilon^3 f_d^2 M^3}{6} \left[ \nu_{s3} + \frac{M-1}{\nu_{i1}-1} (3\nu_{s2}\nu_{i2} + \nu_{s1}\nu_{i3}) + 3 \left( \frac{M-1}{\nu_{i1}-1} \right)^2 \nu_{s1} (1+\alpha) \nu_{i2}^2 \right] \tag{8}$$

where $F_s$ is the spontaneous fission rate, $\varepsilon$ is the detection efficiency, $\nu_{sk}$ and $\nu_{ik}$ are the factorial moments of the spontaneous- and induced-fission multiplicity distributions, and $f_d = e^{-t_{\text{pd}}/\tau} \left( 1 - e^{-t_g/\tau} \right)$ is the doubles gate fraction (predelay $t_{\text{pd}}$, gate width $t_g$, detector die-away $\tau$; LA-13422-M Eq. 5-37). The corresponding triples gate fraction reduces to $f_d^2$ in the narrow-gate R+A scheme used here. For a bare non-multiplying source ($M = 1$, $\alpha = 0$) Eqs. (7)-(9) reduce to $S_{\text{pm}} = F_s \varepsilon \nu_{s1}$, $D_{\text{pm}} = F_s \varepsilon^2 f_d \nu_{s2}/2$, $T_{\text{pm}} = F_s \varepsilon^3 f_d^2 \nu_{s3}/6$ (the c2 cases in this paper); the c3s and 100 g Pu metal cases use $\alpha = 0$ but $M > 1$, and c4s exercises both.

## 3.1. ESARDA Benchmark

The ESARDA (European Safeguards Research and Development Association) Neutron Multiplicity Benchmark [24] is a simulation exercise: given an exact specification of the detector geometry, the sample (mass, density, isotopics), the source rate, and the shift-register parameters, participating codes report $S$, $D$, $T$ for a fixed set of cases (c1, c2-X, c3s/b, c4s/b, …). The detector parameters $\varepsilon$ and the die-away $\tau$ in the spec are anchored to real AWCC measurements at JRC Ispra; the cases themselves are idealised and the "answers" come in two flavours. For non-multiplying $^{252}$Cf cases (c2 family) the validation target is the analytic point-model prediction at the spec $\varepsilon$ and $F_s$. For multiplying or $(\alpha, n)$-active cases (c3s, c4s) the multiplying point model is the first-order target but has known closed-form extensions to higher multiplication [6, 9], and the inter-participant consensus across MCNPX (LANL), MCNP-PTA (JRC) [15], MCNP-PoliMi [1], IPPE, and ONMS [2] (the ESARDA Phase 1+2 participant codes, partly overlapping with the broader tool survey in Table 1) is the de-facto secondary target. We report against both.

### 3.1.1. Detector Geometry

We reproduce the Active Well Coincidence Counter (AWCC) geometry from the ESARDA Neutron Multiplicity Benchmark [24], matching the ONMS implementation [2] for a like-for-like comparison. All geometry parameters are taken from the ONMS GDML (Geant4 Detector Markup Language) model: 42 $^3$He proportional tubes in two rings (21 each at radii 15.40 and 19.05 cm), each with a 50.8 cm active gas region (1.23 cm radius, $5.02 \times 10^{-4}$ g/cm$^3$) plus inactive end regions (steel electronics, dead gas, Al cladding); modelling the full tube housing (1.43 cm outer radius, 60.78 cm length) – even though only the active gas scores detections – ensures that the inactive hardware correctly displaces HDPE rather than being treated as moderator. The tubes sit in a cylindrical HDPE moderator (23.655 cm outer radius, 0.955 g/cm$^3$) enclosed in a 0.2 cm Al shell, with a central sample cavity (11.24 cm radius, 35 cm height) lined by 0.15 cm Al and 0.04 cm Cd (Cd facing the HDPE to absorb returning thermal neutrons). Figure 3 shows a cross-section of the OpenMC model.

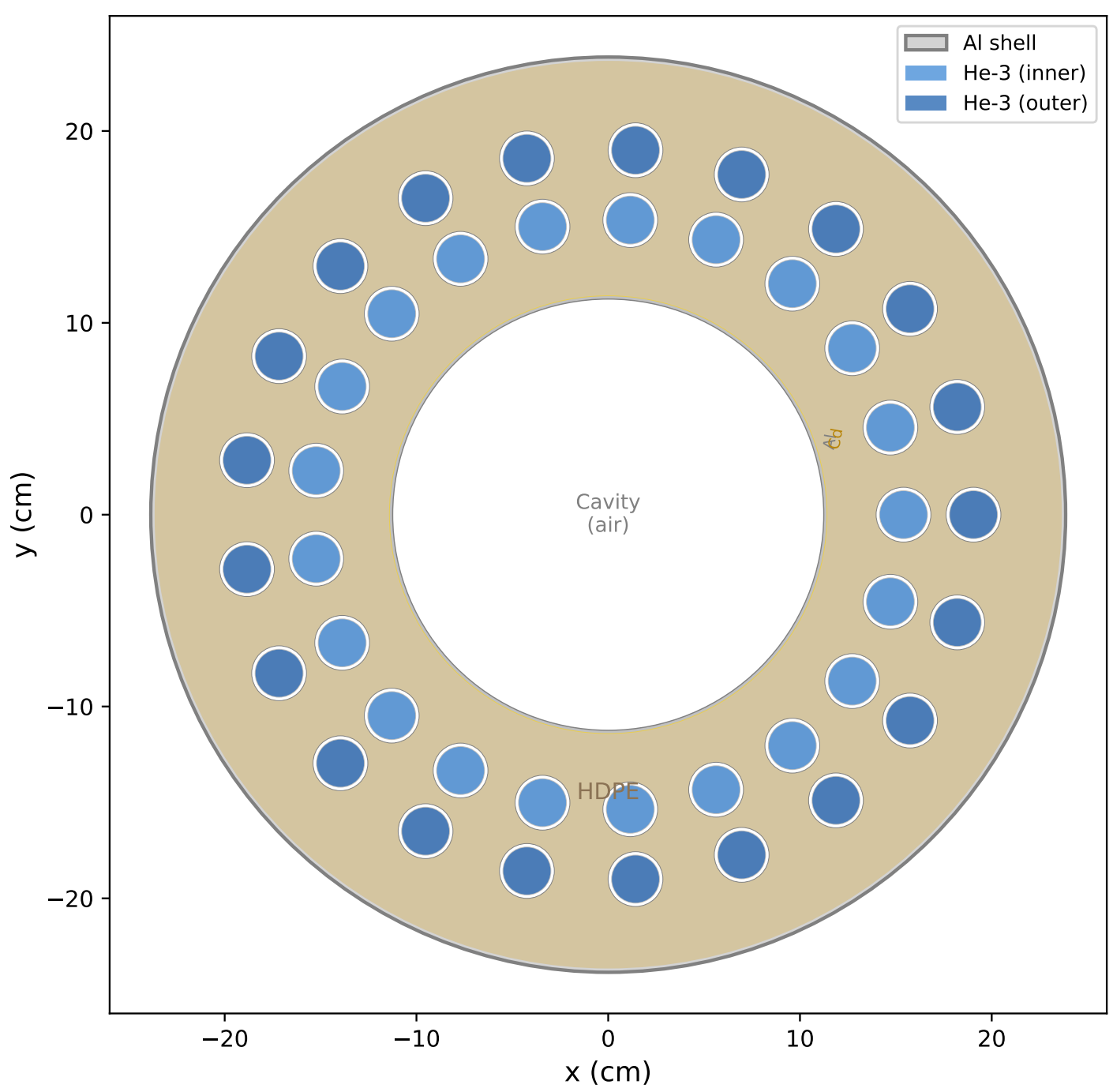

**Figure 3:** Cross-section of the AWCC detector model at $z = 0$. The 42 $^3$He tubes (blue) are arranged in two rings within the HDPE moderator (gray). The central sample cavity (white) is lined with Al and Cd.

The predelay (4.5 $\mu$s) and gate width (64 $\mu$s) match the ESARDA specification [24] Section 3; the long delay is set to 4096 $\mu$s, the value used in LA-13422-M [3] (ESARDA does not specify a value).

#### 3.1.2. Cf-252 Point Source Results

We simulate two ESARDA benchmark cases with $^{252}$Cf point sources at different count rates: c2-10 (8,600 SF/s) and c2-100 (86,000 SF/s). For each case, six independent realisations are run with different random seeds to quantify statistical uncertainties. The simulated $S$, $D$, $T$ are compared to the analytic point-model prediction evaluated at the spec efficiency, source rate, and Holden–Zucker $\nu$ moments (Table 3); running both activities is a cross-rate consistency check that the same code reproduces the same ratios across a factor-of-10 range in count rate, since the point-model ratios are activity-independent invariants of the detector and source physics.

**Two complementary validation modes.** The validation that follows uses two distinct comparisons against the point model, addressing different parts of the simulation chain. The absolute shift-register rates $S$, $D$, $T$ in Table 3 test the full realistic chain end-to-end: source generation, transport, detection, timing, gate-fraction modelling, and shift-register analysis. The per-history factorial-moment ratios in Table 5 bypass the shift register and the gate fraction by tagging each detected neutron with its parent fission, computing the per-fission detection-multiplet factorial moments directly. They isolate transport, detection efficiency, and fission-correlation physics from the die-away/gate-fraction modelling.

The two modes give the same value for the Singles rate but different values for the Doubles and Triples. This follows from the point-model formulas (Eqs. (7)-(9)): $S_{\text{pm}}$ has no gate fraction, while $D_{\text{pm}}$ carries one factor of $f_d$ and $T_{\text{pm}}$ two factors. We therefore report a single $S/S_{\text{pm}}$ comparison (the per-history and shift-register Singles are identical), but two flavours each for Doubles and Triples: shift-register $D/D_{\text{pm}}$, $T/T_{\text{pm}}$ at the simulated $f_d$ in Table 3, and gate-fraction-free $D_{\text{ph}}/D_{\text{pm}}$, $T_{\text{ph}}/T_{\text{pm}}$ in Table 5.

| Source | $\varepsilon_{\text{geom}}$ | $S$ (1/s) | $D$ (1/s) | $T$ (1/s) |
|---|---|---|---|---|
| **c2-10** – $^{252}$Cf, 8 600 SF/s, $M = 1$, $\alpha = 0$<br>PM (ES): [24], Table 2, case 2a | | | | |
| This work | $0.3025 \pm 0.0006$ | $9\,804 \pm 44$ | $3\,000 \pm 102$ | $543 \pm 68$ |
| PM (sim) | 0.3020 | 9 820 | 3 111 | 543 |
| PM (ES) | 0.3095 | 10 000 | 3 251 | 588 |
| **c2-100** – $^{252}$Cf, 86 000 SF/s, $M = 1$, $\alpha = 0$<br>PM (ES): [24], Table 2, case 2b | | | | |
| This work | $0.3015 \pm 0.0004$ | $97\,802 \pm 74$ | $29\,391 \pm 856$ | $4\,421 \pm 1\,445$ |
| PM (sim) | 0.3020 | 97 880 | 31 072 | 5 410 |
| PM (ES) | 0.3095 | 100 000 | 32 509 | 5 885 |
| **c3s** – 10 g Pu metal, $M = 1.12$ (our $k$-eigenvalue; ESARDA spec $M = 1.08$), $\alpha = 0$<br>PM (ES): [24], Table 2, case 3a | | | | |
| This work | $0.296 \pm 0.007$ | $343 \pm 8$ | $80 \pm 8$ | $17 \pm 5$ |
| PM (sim) | 0.296 | 342 | 83 | 17 |
| PM (ES) | 0.316 | 350.4 | 80.8 | 14.0 |
| **Pu metal, ≈ 100 g** – extension beyond ESARDA spec<br>$M = 1.29$ (independent OpenMC $k$-eigenvalue, $k_{\text{eff}} = 0.226$), $\alpha = 0$ | | | | |
| This work | $0.285 \pm 0.001$ | $3\,793 \pm 34$ | $1\,282 \pm 51$ | $554 \pm 65$ |
| PM[7] | 0.285 | 3 795 | 1 310 | 511 |
| **c4s** – 10 g $PuO_2$ powder, $M = 1.014$, $\alpha = 0.853$ (ESARDA spec, see fn.[10])<br>PM (ES): [24], Table 2, case 4a | | | | |
| This work | $0.305 \pm 0.002$ | $1\,978 \pm 16$ | $210 \pm 7$ | $26 \pm 3$ |
| PM (sim) | 0.300 | 1 969 | 199 | 21 |
| PM (ES) | 0.300 | 1 960 | 198.2 | 20.9 |

**Table 3:** Absolute shift-register rates across all ESARDA cases. "This work" rows are means ± standard deviation across $N$ realisations seeded $1, 2, ..., N$ ($N = 6$ for c2-10, c2-100, c4s and the 100 g Pu extension; $N = 9$ for c3s). PM (ES) rows are reproduced from [24] Table 2 (cases 2a, 2b, 3a, 4a for c2-10, c2-100, c3s, c4s); PM (sim) rows are the same point-model expressions evaluated at the simulated $\varepsilon_{\text{geom}}$ for direct comparison with our shift-register output. The 100 g PM row is computed in-house from the multiplying point model [3] at our $k$-eigenvalue $M = 1.29$ and bare-detector efficiency $\varepsilon_{\text{geom}} = \varepsilon_{\text{sim}}/M$ (the two coincide for the non-multiplying c2 cases). Per-history (gate-fraction-free) ratio comparisons against ESARDA participant codes are given separately in Table 5.

The detection efficiency $\varepsilon_{\text{geom}}$ is the probability that a source neutron is captured in a $^3$He volume via $^3\text{He}(n, p)^3\text{H}$, the only reaction we count as a detection. We obtain $\varepsilon_{\text{geom}} = 0.303 \pm 0.001$ on c2-10 (mean over six simulated realisations), compared to the ESARDA-spec $\varepsilon_{\text{geom}} = 0.310$. Two well-understood effects dominate the comparison. First, the inactive tube ends matter: dropping the steel electronics blocks, inactive $^3$He, and Al cladding read from the ONMS GDML and treating the full 60.78 cm tube length as uniform active $^3$He raises $\varepsilon_{\text{geom}}$ by ≈ 4%. Second, after accounting for the geometry effect, a residual offset of ≈ 1–3% remains between our value and the spec; this sits within the "few-percent" inter-participant standard deviations reported for the ESARDA Phase III+IV benchmark [25], and we discuss its likely origin (NJOY processing of the $S(\alpha, \beta)$ bound-atom kernel) in the next paragraph.

The $S(\alpha, \beta)$ thermal-scattering kernel is the bound-atom scattering law for hydrogen in polyethylene. At neutron energies above ≈ 5 eV the scattering off hydrogen is well approximated by free-atom kinematics; at thermal energies the proton is chemically bound in a $CH_2$ chain and its kinematics are dominated by the molecular vibrational modes, so a tabulated $S(\alpha, \beta)$ kernel is required for accurate thermalisation. We invoke this with the OpenMC identifier `c_H_in_CH2`; without it, $\varepsilon_{\text{geom}}$ rises to 0.311, a $\approx 3\%$ effect. OpenMC and MCNP both use $S(\alpha, \beta)$ but process the underlying ENDF data through different NJOY pipelines (OpenMC via direct NJOY/THERMR output, MCNP via its own internally maintained ACE format); these processing differences contribute to the typical inter-code scatter for thermal-well counters [24, 25] alongside differences in the evaluated underlying thermal-scattering data, without being isolated in our comparison.

[7] The 100 g Pu extension lies beyond the ESARDA benchmark range, so no published reference values exist. The PM row is computed in-house from the multiplying point-model expressions [3] evaluated at our $M = 1.29$ from an independent OpenMC $k$-eigenvalue calculation on the bare sample (run_keff.py: $k_{\text{eff}} = 0.226 \pm 0.000$) and the bare-detector efficiency $\varepsilon_{\text{geom}} = \varepsilon_{\text{sim}}/M = 0.285$, with Pu-240 SF moments $(\nu_{s1}, \nu_{s2}, \nu_{s3}) = (2.156, 3.825, 5.336)$, $^{239}$Pu induced moments $(\nu_{i1}, \nu_{i2}, \nu_{i3}) = (3.009, 7.411, 14.593)$, and the spec gate fractions $f_d = 0.6598$, $f_d^2 = 0.4353$ ([24] Table 2 conventions, applied here at our $M$).

For the high-rate c2-100 case, the detection rate of $\approx 10^5$ Hz combined with a 4.5 $\mu$s predelay means that $\approx$47% of consecutive triggers have overlapping gates. The simulation treats $^3$He tubes as ideal absorbers with no dead time, which is appropriate for a simulation benchmark: dead time is a detector electronics effect absent from the idealised Monte Carlo geometry. Any future comparison with experimental data at these rates would need to account for paralyzable or non-paralyzable dead-time models.

Singles and Doubles agree robustly with the point-model expectations: $S$ to better than 0.2% at both count rates, and $D/S$ consistent across the factor-of-10 rate range (0.303 vs. 0.306), demonstrating that the coincidence analysis is rate-stable. Triples are consistent with the point model within the larger third-order statistical uncertainties (relative uncertainty $\approx 13\%$ at c2-10, $\approx 33\%$ at c2-100), as expected for a third-order coincidence rate dominated by accidental-subtraction noise that grows quadratically with rate.

The Rossi-$\alpha$ distribution is the histogram of time differences $\Delta t$ between every ordered pair of detection events. Pairs from the same fission event appear with $\Delta t$ exponentially distributed as $\exp(-\Delta t/\tau_d)$, where $\tau_d$ is the detector die-away time – the time over which a thermalising neutron loses correlation with its fission siblings via absorption in $^3$He or leakage out of the moderator. Pairs from different (uncorrelated) fission events appear uniformly in $\Delta t$ and contribute a flat baseline. The slope of the exponential therefore measures $\tau_d$ directly, and $\tau_d$ in turn sets the gate fraction $f_d = e^{-t_{\text{pd}}/\tau_d}\left(1 - e^{-t_g/\tau_d}\right)$ that converts per-history factorial moments into shift-register rates – the calibration parameter every NMC analysis depends on. A single-exponential plus accidental-baseline fit to the all-pairs Rossi-$\alpha$ distribution from the six combined c2-10 realisations (Figure 4) yields $\tau_{\text{OpenMC}} = 53.9 \pm 1.1\mu s$, $\approx 8\%$ longer than the spec assumption $\tau_{\text{spec}} = 50\mu s$ used in [24]. The corresponding forward gate fractions are $f_d(\tau_{\text{OpenMC}}) = 0.639$ vs $f_d(\tau_{\text{spec}}) = 0.660$, a $\approx 3\%$ offset that propagates directly into the $D/D_{\text{pm}}$ comparison in Table 3. The per-history Doubles ratio (Table 5), which is independent of the die-away shape, gives $D_{\text{ph}}/D_{\text{pm}} = 1.002$.

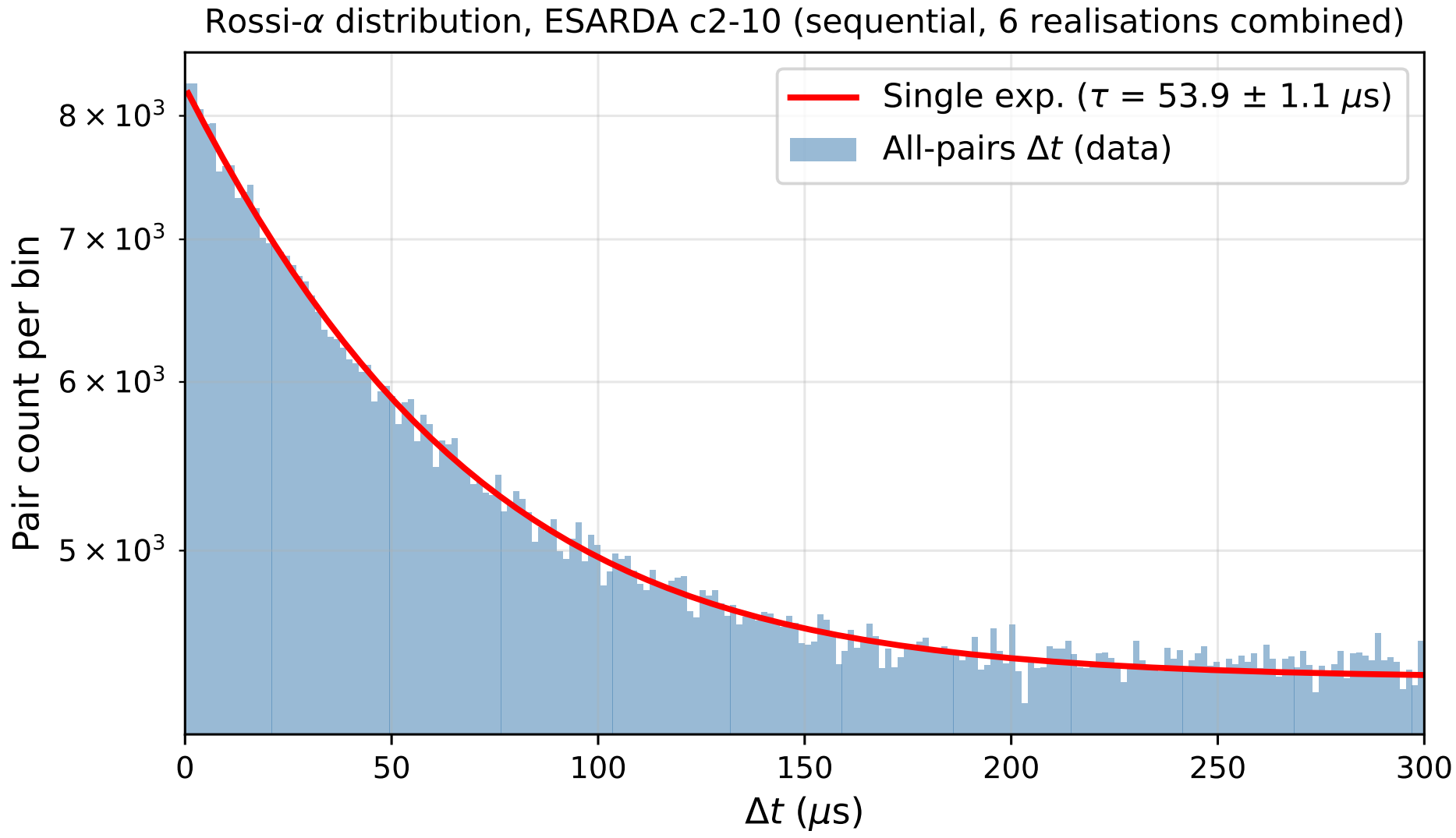

**Figure 4:** Rossi-alpha distribution from the ESARDA c2-10 sequential simulations (six realisations combined, $\approx 2.95 \times 10^5$ detections). Blue: all-pairs time differences within a 300 $\mu$s window. Red: single-exponential plus accidental-baseline fit over 10–200$\mu s$, giving $\tau = 53.9 \pm 1.1 \mu s$.

#### 3.1.3. Cross-Code Validation with ONMS

The preceding ESARDA results validate the full simulation chain (source generation, transport, detection, shift register) as a unit. To isolate and validate the shift register implementation independently of the transport code, we pass two pulse trains to the ONMS pulsetrain analysis library [2]: (1) an ONMS-generated example pulse train of 20,328 Cf-252 detections distributed with the ONMS library, produced by independent Geant4 transport, and (2) an OpenMC-generated pulse train of 52,442 detections from the c2-10 benchmark. ONMS implements its own shift register in C++ with identical parameters (predelay 4.5 $\mu$s, gate 64 $\mu$s, long delay 4096 $\mu$s). Agreement on the ONMS-generated pulse train confirms that the Python shift register is not biased by OpenMC-specific transport artefacts.

| | *ONMS pulse train* | | *OpenMC pulse train* | |
|---|---|---|---|---|
| **Rate** | **Python** | **ONMS** | **Python** | **ONMS** |
| $S$ (1/s) | 2032.8 | 2032.8 | 10553.0 | 10552.9 |
| $D$ (1/s) | 1142.1 | 1143.4 | 4967.0 | 4980.8 |
| $T$ (1/s) | 365.0 | 367.0 | 1774.0 | 1783.8 |

**Table 4:** Cross-code validation of the shift-register implementation, isolated from transport physics. The Python shift register (this work) and the ONMS C++ shift register are applied to two pulse trains: left, an ONMS-generated $^{252}$Cf pulse train (20 328 events) produced by independent Geant4 transport; right, an OpenMC-generated pulse train from the c2-10 benchmark ($\approx 5 \times 10^4$ events). Agreement on the externally generated ONMS pulse train confirms that the Python implementation is not biased by OpenMC-specific artefacts; agreement on both pulse trains is better than 0.6% for all rates.

The agreement is better than 0.1% for Singles and better than 0.6% for Doubles and Triples on both pulse trains. The small residual differences arise from different numerical treatments of gate boundaries.

#### 3.1.4. Comparison with ESARDA Benchmark Participants

We perform two parallel analyses of the same simulated pulse train. The shift-register analysis applies the algorithm of Section 2 to the absolute detection times, yielding $S$, $D$, $T$ that are directly comparable to experimental hardware output and depend on the detector die-away time $\tau$ through the gate fraction $f_d$. The per-history analysis bypasses the shift register: each detected neutron is tagged with the index of the source fission that produced it, and the factorial moments of the per-fission detection-multiplet histogram are computed directly. The per-history Doubles and Triples reduce to $D_{\mathrm{ph}} = F_s \varepsilon^2 \nu_{s2}/2$ and $T_{\mathrm{ph}} = F_s \varepsilon^3 \nu_{s3}/6$, gate-fraction-free expressions that depend only on the bare-detector efficiency and the source-multiplicity moments. This separation lets us validate the transport, detection, and fission-correlation physics independently of any shift-register gate-fraction modelling, and gives the cleanest comparison with the per-history tallies reported by the ESARDA Phase I+II participants [24]. Table 5 collects the per-history ratios for all five validation cases – c2-10, c2-100, c3s, the 100 g Pu extension, and c4s – alongside the ESARDA Phase I+II participant ratios [24] where available.

The per-history analysis on the bare c2-10 case shows excellent agreement (Table 5), consistent with all ESARDA participants to within statistical uncertainties. This supports the correctness of the neutron transport, detection efficiency, and fission-correlation physics in our simulation. Multiplying-sample (c3s, 100 g) per-history ratios are discussed in Section 3.1.4.1, and the $\alpha$-active c4s case in Section 3.1.4.2.

#### *3.1.4.1. Multiplying samples*

To exercise the FREYA induced-fission coupling, we extend the per-history analysis to the ESARDA c3s case (10 g Pu metal, 90% $^{239}$Pu / 10% $^{240}$Pu, $\rho = 19.84$ g/cm$^3$, modelled as a "cubic" cylinder ($H = D$) of radius $\approx 0.43$ cm and height $\approx 0.86$ cm centred in the AWCC cavity; $M = 1.12$ from an independent $k$-eigenvalue calculation, ESARDA spec $M = 1.08$) and to a $\approx 100$ g Pu metal extension (radius $\approx 0.93$ cm, height $\approx 1.86$ cm; $M = 1.29$ from an independent OpenMC $k$-eigenvalue calculation ($k_{\mathrm{eff}} = 0.226$), beyond the ESARDA spec range). For these multiplying $\alpha = 0$ samples, the point-model denominators are evaluated from Eqs. (8)-(9) at the simulated $M$, $\varepsilon_{\mathrm{geom}}$, and the $^{239}$Pu induced-fission moments $\nu_{ik}$; the per-history denominators in Table 5 are the gate-fraction-free forms ($f_d \to 1$, $f_d^2 \to 1$), while the shift-register denominators in Table 3 retain the $f_d$ and $f_d^2$ factors.

The c3s and $\approx 100$ g Pu metal rows of Table 3 give the absolute shift-register rates and the per-history ratios $D_{\mathrm{ph}}/D_{\mathrm{pm}}$ and $T_{\mathrm{ph}}/T_{\mathrm{pm}}$. For c3s, comparing our simulation (at $M = 1.12$ from an independent OpenMC $k$-eigenvalue calculation, $\varepsilon_{\mathrm{geom}} = 0.296$) against the in-house multiplying point model evaluated at the same $M$ and $\varepsilon_{\mathrm{geom}}$: $S/S_{\mathrm{pm}} = 1.004 \pm 0.024$, $D_{\mathrm{ph}}/D_{\mathrm{pm}} = 1.015 \pm 0.086$, $T_{\mathrm{ph}}/T_{\mathrm{pm}} = 1.107 \pm 0.281$. The $T$ deviation reflects our $M = 1.12$ vs spec $M = 1.08$ entering cubically in the Triples expression; the inter-participant scatter on c3s in [24], Table 4 case 3a is itself wide – $T$/PM from 0.99 (MCNPX moments) to 1.40 (IPPE), $D$/PM from 0.98 (MCNPX) to 1.26 (IPPE) – so our $T$ ratio of 1.11 sits well inside that scatter. For the 100 g Pu extension, comparing against the in-house multiplying PM evaluated at our $M = 1.29$ ($k$-eigenvalue) and $\varepsilon_{\mathrm{geom}}$, the $S$ ratio is consistent with unity within statistical uncertainty ($0.999 \pm 0.009$), while $D_{\mathrm{ph}}/D_{\mathrm{pm}} = 1.034 \pm 0.011$ and $T_{\mathrm{ph}}/T_{\mathrm{pm}} = 1.199 \pm 0.050$ deviate by 3–20%, supporting that FREYA reproduces the correct induced-fission coincidence structure through the chain. When evaluating the multiplying point-model denominators $D_{\mathrm{pm}}$ and $T_{\mathrm{pm}}$, we use the bare-detector efficiency $\varepsilon_{\mathrm{geom}} = \varepsilon_{\mathrm{sim}}/M$ rather than the simulated efficiency $\varepsilon_{\mathrm{sim}}$ directly. The reason is that $\varepsilon_{\mathrm{sim}}$ counts every detection, including detections of neutrons born from induced fissions in the sample, so it already carries a factor of $M$ from the multiplication chain; the $M^k$ factors in the closed-form expression account for the same multiplication contribution explicitly, and using $\varepsilon_{\mathrm{sim}}$ in the $\varepsilon^k$ slot would double-count it. The closed-form expression itself is single-generation: it captures only the first round of induced fission, and Böhnel's multi-generation corrections [6] become non-negligible at $M \gtrsim 1.5$. The c3s ($M = 1.12$) and 100 g ($M = 1.29$) cases used here sit well below that threshold, so the closed form is adequate; high-multiplication samples ($M > 1.5$) are reserved for future work.

| Code | $\varepsilon_{\text{geom}}$ | $S/S_{\text{pm}}$ | $D_{\text{ph}}/D_{\text{pm}}$ | $T_{\text{ph}}/T_{\text{pm}}$ |
|---|---|---|---|---|
| **c2-10** – $^{252}$Cf, 8 600 SF/s, $M = 1$, $\alpha = 0$<br>Participant ratios: [24], Table 4, case 2a (point-model denominators from Table 2, case 2a). | | | | |
| PM (ES) | 0.3095 | 1.000 | 1.000 | 1.000 |
| MCNPX (LANL)[8] | 0.3095 | 0.999 | 0.994 | 0.991 |
| MCNP-PTA (JRC) | 0.3095 | 0.998 | 0.996 | 1.074 |
| MCNP-PoliMi | 0.3095 | 1.006 | 1.019 | 1.032 |
| IPPE | 0.3095 | 1.000 | 0.997 | 0.920 |
| MCNPX moments (see fn.8) | 0.3095 | 0.996 | 0.986 | 0.976 |
| ONMS (Geant4)[9] | — | $\approx$ 1.000 | $\approx$ 1.020 | $\approx$ 1.050 |
| This work (per-history) | $0.3025 \pm 0.0006$ | $0.998 \pm 0.005$ | $0.995 \pm 0.007$ | $0.990 \pm 0.010$ |
| **c2-100** – $^{252}$Cf, 86 000 SF/s, $M = 1$, $\alpha = 0$<br>Participant ratios: [24], Table 4, case 2b (point-model denominators from Table 2, case 2b). ONMS values are not reported for c2-100: [2] Figure 6.9 contains sub-panels for c2-10, c3s, and c4s only. | | | | |
| PM (ES) | 0.3095 | 1.000 | 1.000 | 1.000 |
| MCNPX (LANL) | 0.3095 | 0.998 | 0.987 | 0.932 |
| MCNP-PTA (JRC) | 0.3095 | 0.998 | 0.988 | 1.025 |
| MCNP-PoliMi | 0.3095 | 1.006 | 1.020 | 1.103 |
| IPPE | 0.3095 | 1.000 | 0.999 | 0.846 |
| MCNPX moments | 0.3095 | 0.996 | 0.986 | 0.976 |
| This work (per-history) | $0.3015 \pm 0.0004$ | $0.999 \pm 0.001$ | $0.997 \pm 0.003$ | $0.992 \pm 0.005$ |
| **c3s** – 10 g Pu metal, $M = 1.12$ (our $k$-eigenvalue; ESARDA spec $M = 1.08$), $\alpha = 0$<br>Participant ratios: [24], Table 4, case 3a (point-model denominators from Table 2, case 3a). | | | | |
| PM (ES) | 0.316 | 1.000 | 1.000 | 1.000 |
| MCNPX (LANL) | 0.316 | 0.999 | 0.996 | 1.024 |
| MCNP-PTA (JRC) | 0.316 | 0.998 | 1.001 | 1.124 |
| MCNP-PoliMi | 0.316 | 1.002 | 1.013 | 1.052 |
| IPPE | 0.316 | 1.034 | 1.256 | 1.396 |
| MCNPX moments | 0.316 | 0.995 | 0.982 | 0.995 |
| ONMS (Geant4) (see fn.9) | — | $\approx$ 1.000 | $\approx$ 1.040 | $\approx$ 1.220 |
| This work (per-history) | $0.296 \pm 0.007$ | $1.004 \pm 0.024$ | $1.015 \pm 0.086$ | $1.107 \pm 0.281$ |
| **Pu metal, $\approx$ 100 g** – extension beyond ESARDA spec, $M = 1.29$ (independent OpenMC $k$-eigenvalue), $\alpha = 0$<br>No ESARDA participant ratios (case is outside the benchmark range); in-house point-model denominator follows [3], see footnote in Table 3. | | | | |
| PM (in-house, see fn.[7]) | 0.285 | 1.000 | 1.000 | 1.000 |
| This work (per-history) | $0.285 \pm 0.001$ | $0.999 \pm 0.009$ | $1.034 \pm 0.011$ | $1.199 \pm 0.050$ |
| **c4s** – 10 g $PuO_2$, $M = 1.014$, spec $\alpha = 0.853$[10]<br>Participant ratios: [24], Table 4, case 4a (point-model denominators from Table 2, case 4a). | | | | |
| PM (ES) | 0.300 | 1.000 | 1.000 | 1.000 |
| MCNPX (LANL) | 0.300 | 1.051 | 1.102 | 1.167 |
| MCNP-PTA (JRC) | 0.300 | 1.021 | 1.103 | 1.269 |
| MCNP-PoliMi | 0.300 | 1.022 | 1.105 | 1.162 |
| IPPE | 0.300 | 1.082 | 1.318 | 1.357 |
| MCNPX moments | 0.300 | 1.047 | 1.079 | 1.107 |
| ONMS (Geant4) (see fn.9) | — | $\approx$ 1.020 | $\approx$ 1.100 | $\approx$ 1.250 |
| This work (per-history) | $0.305 \pm 0.002$ | $1.005 \pm 0.008$ | $1.068 \pm 0.033$ | $1.206 \pm 0.092$ |

**Table 5:** Per-history (gate-fraction-free) multiplicity results normalised to the point model, for all five validation cases. Participant ratios are reproduced from [24] Tables 3 and 4 (cases 2a, 2b, 3a, 4a for c2-10, c2-100, c3s, c4s); the 100 g Pu extension lies outside the ESARDA benchmark range and shows only the in-house PM denominator. The MCNP-PoliMi entries are computed from raw rates in [24] Table 3 divided by [24] Table 2 point-model values; ONMS entries (only for cases that appear in Kütt's Figure 6.9 sub-panels) are read off the bar charts as documented in the c2-10 footnote.

#### *3.1.4.2. Plutonium-oxide sample with* $(\alpha, n)$*: case c4s*

To exercise the full pipeline including the $(\alpha, n)$-source contribution, we run the ESARDA c4s case: 10 g of $PuO_2$ powder (Pu mass fractions 2% $^{238}$Pu / 60% $^{239}$Pu / 25% $^{240}$Pu / 8% $^{241}$Pu / 5% $^{242}$Pu, density 2 g/cm$^3$) modelled as a "cubic" cylinder of radius $\approx 0.93$ cm and height $\approx 1.85$ cm centred in the AWCC cavity. ALPHANSO [20], run on the same composition with modern $(\alpha, n)$ cross-section data, returns $\alpha = 0.78$. The ESARDA spec value, computed by Peerani *et al.* in 2006 with SOURCES-4C / JENDL/A-2003, is $\alpha = 0.853$ – $\approx 9\%$ higher; the difference is a real cross-section-data discrepancy on the dominant $^{17,18}\mathrm{O}(\alpha, n)$ targets, not a bug in either tool (see[10]). To enable an apples-to-apples comparison with the ESARDA participants, we re-ran the c4s case with the $(\alpha, n)$ rate scaled to the spec $\alpha = 0.853$ over six independent realisations of 10 s each (FREYA-enabled induced fission, single-thread for FREYA serialisation).

The c4s rows of Table 3 and Table 5 summarise the result: per-history $S/S_{\mathrm{pm}} = 1.005 \pm 0.008$, $D_{\mathrm{ph}}/D_{\mathrm{pm}} = 1.068 \pm 0.033$, $T_{\mathrm{ph}}/T_{\mathrm{pm}} = 1.206 \pm 0.092$, all well inside the ESARDA participant scatter for case 4a ([24] Tables 3-4), with the $S$ ratio closer to the point model than any of the four 2006 participants.

## 3.2. Sources of uncertainty and model sensitivity

Table 6 summarises the dominant contributions to the validation uncertainties, the treatment used in this work, and the magnitude of each effect where it is quantified in the manuscript.

| **Source** | **Affected quantities** | **Treatment** | **Magnitude / comment** |
|---|---|---|---|
| Monte Carlo counting statistics | $S$, $D$, $T$ (esp. $T$) | Six to nine independent realisations per case | Per-realisation rel. unc.: $S < 0.5\%$, $D \approx 3\%$, $T \approx 13\%$ (c2-10) – 33% (c2-100). $T$ dominates. |
| Detection efficiency / geometry | $S$, $D$, $T$ normalisation | AWCC dimensions read from ONMS GDML; inactive tube ends, steel electronics, Al cladding included | $\varepsilon_{\mathrm{geom}} = 0.303 \pm 0.001$ vs. spec 0.310 ($\approx 2\%$); dropping inactive ends raises $\varepsilon_{\mathrm{geom}}$ by $\approx 4\%$. |
| Thermal scattering $S(\alpha, \beta)$ in polyethylene | Moderation, $\varepsilon$ | `c_H_in_CH2` bound-atom kernel in OpenMC | Without it, $\varepsilon_{\mathrm{geom}} \to 0.311$ ($\approx 3\%$ shift). Residual code-to-code offset is within the few-percent inter-participant scatter reported in [25]. |

[8]"MCNPX (LANL)" and "MCNPX moments" are both LANL submissions to the ESARDA Phase I+II benchmark using the MCNPX transport code, but with different shift-register analysis methods: the former post-processes an MCNPX-generated pulse train through an external shift-register analyser, while the latter uses MCNPX's direct multiplicity-moment tallies (available only for zero-dead-time cases). See [24] Section 5.1 for the methodology breakdown.

[9]ONMS S/D/T values for c2-10, c3s, and c4s are read off the bar charts in Kütt's 2016 PhD thesis [2], Figure 6.9 (epI+II-c2-10, epI+II-c3s and epI+II-c4s sub-panels) (Phase I+II ESARDA Benchmark results), and are precise to the chart's resolution ($\approx \pm 0.01$ for $S$ and $D$, $\approx \pm 0.03$ for $T$). Figure 6.9 reports only the S/D/T ratios normalised to the point model and does not give the bare-detector efficiency, so the $\varepsilon_{\mathrm{geom}}$ entry for ONMS is left blank. ONMS results are not part of the Peerani 2006 benchmark exercise (ONMS was first published in 2013).

[10]The ALPHANSO–ESARDA $\alpha$ discrepancy (native $\alpha = 0.78$ vs. spec $\alpha = 0.853$) reflects shifts of 5–10% in the evaluated $^{17,18}\mathrm{O}(\alpha, n)$ cross sections across the JENDL/A → JEFF-3.3 → TENDL-2021 lineage; independent inter-code benchmarks favour the lower (newer-data) values. The c4s rates reported here use the rescaled $\alpha = 0.853$; using the native $\alpha = 0.78$ would shift the absolute Singles by $\approx -9\%$ without changing the validation conclusions.

| Source | Affected quantities | Treatment | Magnitude / comment |
|---|---|---|---|
| FREYA multiplicity moments | $D$, $T$ (sampled $P(\nu)$ accuracy) | Direct comparison to Holden–Zucker [22] for $^{252}$Cf only (Figure 2); $^{240}$Pu SF and $^{239}$Pu induced fission rely on FREYA's fragment-evaporation framework without a tabulated $P(\nu)$ cross-check. | $^{252}$Cf: $\nu_{s2}$ within 0.1%, $\nu_{s3}$ within 0.6%. Floor/ ceil two-point baseline is $-12\%$ on $\nu_{s2}$, $-38\%$ on $\nu_{s3}$ for context. |
| ALPHANSO $(\alpha, n)$ cross-section libraries | c4s source composition; $S$, $D$, $T$ for c4s | ESARDA-spec $\alpha = 0.853$ used in the c4s validation; native ALPHANSO prediction $\alpha = 0.78$ documented in[10] | Modern-library vs. JENDL/A-2003 spec differ by $\approx 9\%$ on the $(\alpha, n)$-source rate; absorbed into the $\alpha$ rescaling for the participant comparison. |
| Detector die-away $\tau$ / gate fraction $f_d$ | Shift-register $D$, $T$ | $\tau$ fitted from c2-10 Rossi-$\alpha$ distribution; per-history $D$ and $T$ reported in parallel and are gate-fraction-free | Fitted $\tau = 53.9 \pm 1.1\mu s$ (spec $50\mu s$); $f_d^{\text{sim}} = 0.639$ (spec 0.660, $\approx 3\%$ diff). |
| Shift-register boundary conventions | Small differences in $D$, $T$ | Cross-validation against ONMS C++ shift register on independent and shared pulse trains (Table 4) | Sub-0.1% on $S$, $< 0.6\%$ on $D$ and $T$. |
| `FileSource` sampling mode | Correlated $D$, $T$ if uncorrected | Sequential-iteration `FileSource` mode used throughout (see Section 2.4) | Default with-replacement sampling inflates the c2-10 Doubles by $\approx 31\%$ for $^{252}$Cf the sequential mode eliminates the bias by construction. |

**Table 6:** Main sources of uncertainty and model sensitivity in the validation calculations. "Magnitude / comment" entries quote numbers established elsewhere in the manuscript; entries without a numerical magnitude are cross-validated rather than independently estimated.

# 4. Validation Scope and Known Limitations

The present study validates the framework as a simulation-benchmark tool rather than as a complete experimental assay model. Direct comparison with measured Pu-bearing items requires detector-specific dead-time and electronics models and access-controlled reference data, and is therefore left to future work. We summarise here the limits of the present validation and the implementation-scope caveats that affect how the framework should be applied beyond the cases in this paper.

**Not covered.** The present validation does not include: experimental measurements on Pu-bearing items; high-multiplication samples ($M > 1.5$), where Böhnel-style multi-generation corrections [6] become essential in the analytic comparison baseline (the Monte Carlo simulation itself has no such limit); detector and electronics dead-time effects; MPI-parallel sequential `FileSource` iteration; a head-to-head comparison between FREYA and CGMF [23]; and validation on multi-region source distributions (per-event rejection against arbitrary `openmc.Cell` / `Region` unions ships in the framework but is exercised only on cylindrical samples here).

**Implementation-scope caveats.** Three caveats are relevant for users who plan to scale the framework beyond the cases reported here. First, dead time is absent by design in the ESARDA Monte Carlo benchmark and in the present implementation: the $^{3}$He tubes are treated as ideal absorbers with zero dead time, which is appropriate for benchmarking but must be replaced with paralyzable or non-paralyzable models for high-rate experimental data. Second, FREYA's Fortran state is non-re-entrant and is therefore serialised behind an OpenMP critical section; this limits OpenMP scaling on multiplying samples (measured 1.1–1.2 $\times$ at 8 threads on c3s, against 2.7 $\times$ for non-multiplying c2-10), and the recommended scale-out is over independent realisations rather than threads when multiplication is significant. Third, the sequential-`FileSource` mode in the provided OpenMC fork supports OpenMP multi-thread parallelism but not MPI multi-rank distribution; the natural MPI extension partitions the source bank into rank-local contiguous slices and is not exercised here.

# 5. Conclusion

We presented PyNMC, an open-source, Python-scriptable framework for neutron multiplicity counting simulation that couples OpenMC for transport, FREYA for event-by-event correlated prompt-fission neutron emission, and ALPHANSO for native $(\alpha, n)$-source estimates, together with collision-level time-tagged event recording and a Python shift-register post-processor cross-validated against ONMS. As a by-product we implemented a sequential-iteration mode in OpenMC's `FileSource` (in the provided OpenMC fork; upstream submission planned) that fixes a sampling-with-replacement bias affecting time-correlated calculations.

Validation against the ESARDA Neutron Multiplicity Benchmark covers bare $^{252}$Cf (c2-10, c2-100), the c3s low-multiplication Pu metal case, and the $(\alpha, n)$-active $PuO_2$ case c4s (spec $\alpha = 0.853$); an internal $\approx 100$ g Pu metal stress test extends the multiplying-metal study to $M = 1.29$ and lies outside the ESARDA participant range. The Singles rate agrees with point-model predictions to better than 0.1% on the bare cases and 0.5% on c4s, the per-history Doubles ratio sits within a few percent of unity across all five cases (Table 5), and the Python shift register agrees with the ONMS C++ implementation to better than 0.6% on all three rates.

The present validation establishes the framework for ESARDA-style simulation benchmarks and low-multiplication NMC studies; the validation scope and forward-looking extensions are detailed in Section 4 and Section 5.1. Built on a widely used open-source Monte Carlo transport code and shipped with a Dockerfile that pins the main scientific dependencies and records the validation toolchain, the framework offers a transparent baseline for verification-technology development and benchmark-driven detector studies.

## 5.1. Outlook

The detector geometry is built from dimensions extracted from the ONMS GDML file `awcc.gdml`, with no AWCC-specific assumptions baked into the source, transport, or shift-register layers. Other thermal-well counters used in safeguards practice (HLNCC, PSMC) share the same architectural elements (HDPE moderator, $^{3}$He tubes, optional Cd liner) and require only a geometry-level change to be supported. Extending validation to PSMC against measured Pu data is the natural next step in that direction; the remaining items not covered by the present validation are listed in Section 4.

# 6. Program Summary

**Program title.** `nmc` (PyNMC); the repository, import path, and Docker image are all named `nmc`, while PyNMC is the display name used in the README and citation file.

**Licensing provisions.** MIT License.

**Distribution archive.** `github.com/cfichtlscherer/nmc` (tagged release `v1.0`).

**Programming language.** Python 3.10+, C++ (OpenMC modifications).

**Build-time toolchain (as installed in the validation image).** Ubuntu 24.04.4 noble; gcc / gfortran 13.3.0 (apt `13.3.0-6ubuntu2~24.04.1`); cmake 3.28.3 (apt `3.28.3-1build7`); CPython 3.12.3 (apt `3.12.3-0ubuntu2.1`); OpenMC fork commit `f9073e33d` reports as `0.15.3-dev254`. The Dockerfile does not pin these via apt version specifiers, so a fresh `docker build` against current Ubuntu 24.04 archives may pick up any later point release. Statistical reproducibility (within Monte Carlo scatter) holds across rebuilds; bit-identical reproduction is not claimed (see footnote[5]).

**Dependencies.** OpenMC built from `cfichtlscherer/openmc nmc` branch at commit `f9073e33d` (carries the sequential-`FileSource` and `sample_nuclide` rounding-tolerance patches; built with `OPENMC_USE_FREYA`); FREYA 2.0.3 vendored as a prebuilt binary at `vendor/freya2.0.3/lib/libFission.so` (mirrored at `github.com/cfichtlscherer/freya2.0.3` commit `5f18954`; LLNL UCRL-CODE-224807, BSD-3); ALPHANSO 1.0.1 (PyPI release); NumPy, h5py (versions follow the OpenMC requirements). The Dockerfile pins all of these.

**Nuclear data.** ENDF/B-VIII.0 multi-temperature HDF5 library distributed by the OpenMC project (anl.box.com Box URL pinned in the Dockerfile, ≈ 7 GB compressed / ≈ 43 GB extracted, 250–2500 K). Downloaded automatically on first build; the thermal-scattering tables are then filtered to 294 K (all 111 SAB nuclides retained, single temperature each), bringing the in-image cross-section directory to ≈ 9 GB. Neutron data files are kept multi-temperature. The full multi-temperature thermal library can be retained with `--build-arg NMC_FILTER_XS=0`; bind-mounting a local library at `/opt/nmc/xs` is also supported.

**Memory requirements.** ≈ 2 GB resident for c2-10 / c4s; ≈ 4–6 GB peak for c2-100 (large source bank). Build host: ≈ 8 GB free RAM and ≈ 60 GB free disk for the multi-temperature image.

**Nature of problem.** Simulation of neutron multiplicity counter response for nuclear safeguards and arms control verification. Requires correct fission multiplicity correlations, time-correlated particle transport, and shift register analysis.

**Solution method.** Monte Carlo particle transport (OpenMC) with event-by-event correlated prompt-fission emission (FREYA), $(\alpha, n)$-source terms (ALPHANSO), collision-level time-tagged event recording, and Python shift register post-processing.

**Smoke-test expected output.** `docker run --rm nmc` runs `examples/01_cf252_in_awcc.py` and prints, for one c2-10 realisation: $S \approx 9\,800$ s$^{-1}$, $D \approx 2\,950$ s$^{-1}$, $D/S \approx 0.30$, $\varepsilon \approx 0.302$. Statistical scatter across realisations is ≈ 0.5% on $S$, ≈ 3% on $D$, ≈ 13% on $T$.

**Build time.** Clean (no-cache) `docker build` on the reference host (see footnote[2]): 1 h 03 min, dominated by OpenMC compilation and the ≈ 7 GB cross-section download.

**Running time.** On the reference host (see footnote[2]): smoke test (default `docker run`, one c2-10 realisation) 14 s; isolated single c2-100 realisation 4 min 03 s; isolated single c4s realisation 19 s; bare-source c2 validation campaign (12 realisations) 27 min 00 s; full paper validation suite (33 realisations across all five cases) 30 min 56 s. All validation scripts run single-threaded for determinism (see footnote[3]).

**Limitations.** No electronics dead-time model (ideal absorbers); no experimental data validation; no multi-region / spatially non-uniform source distributions exercised; no high-multiplication validation beyond $M \approx 1.3$; MPI-parallel sequential `FileSource` not implemented (OpenMP only); FREYA's Fortran state is non-re-entrant and is serialised behind an OpenMP critical section, limiting strong scaling on multiplying samples (see Section 4 for the full list).

## 6.1. Reproducibility and User Workflow

**Repository and license.** The source code, simulation scripts, and a Dockerfile providing containerised builds with pinned upstream dependencies are available at `github.com/`

`cfichtlscherer/nmc` under the MIT license. The released version associated with this paper is tagged on GitHub at the commit listed below, rather than left on a mutable branch.

**Workspace layout.** Repository organisation distinguishes minimal reproducers from full validation runs. `nmc/examples/` contains single-realisation scripts that print to stdout in seconds and serve as tutorials and CI smoke tests; `nmc/validation/` contains the multi-realisation campaigns that write the JSON outputs feeding the paper tables; the `Dockerfile` wires these against pinned versions of the upstream tools (see below) and downloads the cross-section data on first build. Each script's docstring names the paper section, table, or figure it reproduces, and `nmc/examples/README.md` indexes the mapping.

**One-command reproduction.** The full reproducibility chain from a clean machine is

```
git clone https://github.com/cfichtlscherer/nmc.git && cd nmc
docker build -t nmc . && docker run --rm nmc
```

which builds the image (including the multi-temperature ENDF/B-VIII.0 cross-section download) and reproduces a single c2-10 realisation. The pre-built ENDF/B-VIII.0 bundle published by the OpenMC project ships six temperatures (250–2500 K, ≈ 43 GB extracted) per nuclide, dominated by thermal-scattering tables at one-degree spacing (e.g. `c_H_in_H2O` carries 94 temperatures from 273 K to 950 K). For NMC at room temperature only the 294 K data are exercised, so by default the Dockerfile post-processes the extracted thermal-scattering directory in place, retaining only the 294 K group of every `thermal/*.h5` file (all 111 thermal SAB tables and all 768 nuclides preserved, one temperature each). This reduces the in-image cross-section directory from ≈ 43 GB to ≈ 9 GB and the resulting image to ≈ 14 GB. The neutron-data files (`neutron/*.h5`, ≈ 5 GB) are kept multi-temperature because their resonance-region structures resist a clean h5py-level temperature filter. Users who require the full multi-temperature thermal library can opt out with `docker build --build-arg NMC_FILTER_XS=0 -t nmc .`, recovering the ≈ 50 GB image. Per-script invocation inside the image is

```
docker run --rm nmc python /opt/nmc/<path>
```

Specifically, `examples/01_cf252_in_awcc.py` reproduces a single c2-10 realisation; `examples/03_pnu_synthetic.py` and `examples/04_pnu_figure.py` reproduce Table 2 and Figure 2; `examples/05_awcc_geometry_figure.py` reproduces Figure 3 from the geometry parameters alone (no transport required); `examples/06_rossi_alpha_figure.py` reproduces Figure 4 from the c2-10 collision-track output; `validation/run_validation.py` produces the c2-10 / c2-100 rows of Table 3; `validation/run_pu_metal.py` produces the c3s and 100 g Pu metal rows; and `validation/run_pu_oxide.py` produces the c4s rows of Table 3 and Table 5.

**Pinned versions and commit hashes.** The Docker image fixes the following exact versions: Python 3.10; FREYA 2.0.3 vendored as a prebuilt binary[11] in the framework repository at `vendor/freya2.0.3/` (LLNL UCRL-CODE-224807, BSD-3; source preserved alongside the binary, mirror at `github.com/cfichtlscherer/freya2.0.3` commit `5f18954`); ALPHANSO 1.0.1 (PyPI release `alphanso==1.0.1`, source at `github.com/cfichtlscherer/alphanso`); OpenMC built from the `nmc` branch of `github.com/cfichtlscherer/openmc` at commit `f9073e33d` (carries the sequential-`FileSource` and `sample_nuclide` rounding-tolerance patches described in Section 2); and the ENDF/B-VIII.0 multi-temperature cross-section release distributed by the OpenMC project. The framework repository itself is `github.com/cfichtlscherer/nmc` at tagged release `v1.0` for the version associated with this manuscript. NumPy and h5py versions follow the OpenMC requirements at the pinned commit.

[11]FREYA's Fortran-side `RANDOM_NUMBER` intrinsic is implemented by the linked `libgfortran` at compile time, and different `gfortran` releases ship different default RNG algorithms (KISS64 in `gcc` ≤ 6, xoshiro256** in `gcc` ≥ 7). Because the higher factorial moments $\nu_{s2}$ and $\nu_{s3}$ are dominated by the rare high-multiplicity tail of $P(\nu)$ (e.g. $P(\nu = 8) \approx 0.002$ for $^{252}$Cf), they are sensitive to the specific RNG stream at finite sample size. To make the simulation reproducible host-independently we ship a statically-linked FREYA binary in `vendor/freya2.0.3/`; the LLNL source distribution and license files are preserved alongside it for users who wish to rebuild against their own toolchain.

Every paper-table row and figure therefore has a single script in the image that produces it, and every dependency the script needs is fixed by the image build.

# 7. Declaration of Generative AI Use

During the preparation of this work, the author used Claude (Anthropic, Opus 4.6 and Opus 4.7) to assist with creating simulation input files, post-processing scripts, formatting figures, and editing text. After using this tool, the author reviewed and edited all content as needed and takes full responsibility for the content of the published article.